\documentclass{report}
\usepackage{epsfig}

\newcommand{\be}{\begin{equation}}
\newcommand{\ee}{\end{equation}}

\renewcommand{\abstract}[1]{{ \footnotesize \noindent {\bf Abstract} #1 \\}}
\renewcommand{\author}[1]{\subsubsection*{#1}}
\newcommand{\address}[1]{\subsubsection*{\it#1}}

\newcommand{\ARAA}{ARA\&A}

\newcommand{\AaA}{A\&A}

\newcommand{\ApJ}{ApJ}

\begin{document}

\chapter*{CMB and Cosmological Parameters: Current Status
and Prospects.}

\author{Alessandro ~Melchiorri$^{1}$}

\address{$^{1}$ Denys Wilkinson Building,
University of Oxford, Keble Road, Oxford, OX1 3RH, UK.}

\abstract{The last years have been an exciting period for
the field of the Cosmic Microwave Background (CMB) research.
With recent CMB balloon-borne and ground-based experiments we are
entering a new era of 'precision' cosmology that enables us to use
the CMB anisotropy measurements to constrain the cosmological parameters
and test new theoretical scenarios.}


\section{Introduction}

The last years have been an exciting period for
the field of the Cosmic Microwave Background (CMB) research.
With recent CMB balloon-borne and ground-based experiments we are 
entering a new era of 'precision' cosmology that enables us to use 
the CMB anisotropy measurements to constrain the cosmological parameters 
and the underlying theoretical models.

Coeherent oscillations in the Cosmic Microwave Background
anisotropies angular power spectrum 
have been predicted since long time from simple assumptions about 
scale invariance and linear perturbation theory
(see e.g., \cite{Peeb1970}, \cite{SZ70}, \cite{wilson}, \cite{vittorio},
\cite{bondefstathiou}). The physics of these oscillations and their 
dependence on the various cosmological parameters has been described in 
great detail in many reviews (\cite{review}, \cite{review2}, 
\cite{review3}, \cite{review4}, \cite{review5}, \cite{review6}). 
Basically, on sub-horizon scales, prior to recombination, photons and 
baryons form a tightly coupled fluid that performs acoustic
oscillations driven by the gravitational potential.
These acoustic oscillations define a structure of peaks in
the CMB angular power spectrum that can be measured today.

With the TOCO$-97/98$ (\cite{torbet},\cite{miller}) 
and Boomerang-$97$ (\cite{mauskopf}) experiments a firm detection of
a first peak on about degree scales has been obtained. 
In the framework of adiabatic Cold Dark Matter (CDM) models, the
position, amplitude and width of this peak provide strong supporting 
evidence for the inflationary predictions of
a low curvature (flat) universe and a scale-invariant primordial 
spectrum (\cite{knox}, \cite{melchiorri}, \cite{tegb97}).

The new experimental data from BOOMERANG LDB (\cite{netterfield}), 
DASI (\cite{halverson}) and MAXIMA (\cite{lee})
have provided further evidence for the presence of the first peak
and refined the data at larger multipole. 
The combined data clearly suggest the presence of 
a second and third peak in the spectrum, confirming the model
prediction of acoustic oscillations in the primeval plasma 
and sheding new light on various cosmological and 
inflationary parameters (\cite{debe01}, \cite{wang}, \cite{pryke}).

\section{The Current Observational Status.}

\begin{figure}[htb]
\begin{center}
\includegraphics[angle=-90,scale=0.45]{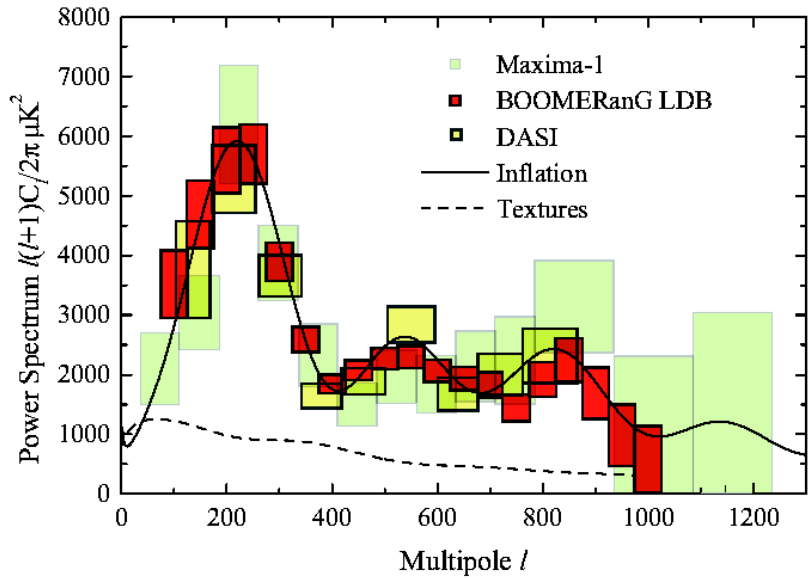}
\end{center}
\caption{BOOMERanG, DASI and MAXIMA data togheter with an inflationary model
and a global textures model.}
\label{fig1}
\end{figure} 

On April 30th 2001, at the same time, $3$
different teams,  BOOMERanG \cite{netterfield}, DASI \cite{halverson} 
and MAXIMA \cite{lee} reported a detection of multiple features in 
the CMB angular power spectrum. 

{\bf The BOOMERanG experiment.} BOOMERanG is a scanning balloon 
experiment aimed at producing accurate 
and high signal/noise maps of the CMB sky and constraining the power
spectrum in the $50 < \ell < 1000$ range. 
The BOOMERanG experiment has been described in 
\cite{piacentini} and \cite{debe99}. 
All the relevant informations about the collaboration can be found
in the 'official' websites: {\it \bf http://oberon.roma1.infn.it/boom} and 
{\it \bf http://www.physics.ucsb.edu/\~boomerang/}.

The BOOMERanG group carried out a long duration flight 
(December 1998/ January 1999) called the Antarctica or
LDB flight. Before this, there was a 'test flight' on North
America from which the first power spectrum results were
released (\cite{mauskopf}, \cite{melchiorri}, \cite{piacentini}).
From the test flight a $\sim 4000$ $16'$ pixel map 
at $150 GHz$ produced a firm detection of 
a first peak in the CMB angular power spectrum.

For the antarctica flight, coverage of $4$ frequencies with $16$
bolometers in total were available. 
BOOMERanG LDB measured $8$ pixels in the sky simultaneously . 
Four pixels feature multiband photometers ($150$, $240$ and
$410$ GHz), two pixels have single-mode, diffraction limited
detectors at $150$ GHz and two pixels have single-mode,
diffraction limited detectors at $90$ GHz.
The NEP of these detectors is below $200 \mu K_{CMB}\sqrt{s}$ at 
$90$, $150$ and $240$ GHz and the angular resolution ranges
from $12$ to $18$ arcminFWHM.

The istrument was flown aboard a stratospheric balloon at
$38 Km$ of altitude to avoid the bulk of atmospheric emission
and noise.
During a long duration balloon flight of $\sim 11$ days 
carried out by NASA-NSBF around Antarctica in $1999$, 
BOOMERanG mapped $\sim 1800$ square degrees in a region of the 
sky with minimal contamination from the galaxy.

The most recent analysis of the BOOMERanG data has been presented 
in \cite{netterfield}. The observations taken from $4$ 
detectors at $150$ GHz in a dust-free ellipsoid central region 
of the map ($1.8 \%$ of the sky) have been analyzed using the
methods of (\cite{borrill}, \cite{hivon}, \cite{prunet}).
The gain calibration are obtained from observations of the CMB dipole.

The CMB angular power spectrum, estimated in $19$ bands
centered between $\ell=50$ to $\ell=1000$ is shown in Figure 1.
The error bars on the $y$ axis are correlated at about $\sim 10 \%$.
A first peak is clearly evident at $\ell \sim 200$ and 
$2$ subsequent peaks can be see in the figure.
Not shown in the figure is an additional $10 \%$
calibration error (in $\Delta T$) and the uncertainty in the beam 
size ($12.9' \pm 1.4'$).

Calibration error does not affect the shape of the power spectrum,
producing only an overall shift in amplitude.

On the contrary, since the beam resolution affects the 
$C_\ell$ spectrum, a small uncertainty 
$\Delta \sigma^2_{beam}=\sigma^2_{beam}-(\sigma_{beam}')^2$ 
in the telescope beam produce a correlated $\ell$-dependent 
'calibration error' of $\sim (1+\ell^2\Delta\sigma^2_{beam})C_\ell$
(\cite{bridle})

The beam uncertainty can change the relative amplitude
of the peaks, but cannot introduce features in the spectrum.

{\bf The DASI experiment.}  The DASI experiment is a ground based 
compact interferometer constructed specifically for observations of the CMB.
A description of the instrument can be found in 
\cite{halverson} and \cite{leitch}.
and all the relevant information about the team can be
obtained from the DASI website:{\it \bf http://astro.uchicago.edu/dasi/}.

The specific advantage of interferometers is in reducing the
effects of atmospheric emission \cite{lay}.
DASI is composed of $13$ element interferometers
with correlator operating from $26$ to $36$ GHz.
The baseline of DASI cover angular scales from $15'$ to
$1.4^o$.

Interferometry is a technique that differs in many
fundamental ways from those used by BOOMERanG and other
map-making CMB experiments.
Interferometers directly sample the Fourier
transform of the sky brightness distribution and 
the CMB power spectrum can be computed without
going through the map making process.
In this sense, the DASI result provides a real
independent observation of the CMB angular spectrum.

The most recent analysis of the DASI data has been 
presented in \cite{halverson}. 
The observations have been taken over $97$ 
days from the South-Pole during the austral summer
at frequencies between  $26$ and $36$ GHz.
The calibration was obtained using bright astronomical
sources.

The CMB angular power spectrum estimated in $9$ bands
between $\ell=100$ to $\ell=900$ is also shown in Figure 1.
There is a $\sim 20 \%$ correlation between the data points.
Not shown in the figure is an $\sim 8\%$ calibration error,
while the beam error is negligible.
The DASI team found no evidence for foregrounds other than
point sources (which are the dominant foregrounds at those
frequencies (see e.g. \cite{efstteg}, \cite{tegfor})).
Nearly $30$ point sources have been detected in the DASI
data while a statistical correction has been made for residual
point sources that were too faint to be detected.

{\bf The MAXIMA experiment.} MAXIMA-I is another balloon experiment, 
similar in many aspects to BOOMERang but not long-duration.
A description of the instrument can be found in \cite{lee}
and all the relevant informations about the team can be
obtained from the MAXIMA website:
{\it \bf http://cosmology.berkeley.edu/group/cmb/}.       
In the latest analysis (\cite{lee}) the data from 
$3$, $150$, GHz very sensitive bolometers has been analyzed in order to
produce a $3'$ pixelized map of about $10$ by $10$ degrees. 
The previous analysis of about the same data\footnote{The $240$ GHz channel 
has been excluded because it did not pass consistency
tests above $\ell=785$.} based on a $5'$ pixelization
(\cite{hanany}) has therefore been extended to $\ell=1235$.
The map-making method used by the MAXIMA team is extensively discussed
in \cite{stompor2}. The data are calibrated using the CMB dipole.

The MAXIMA-I datapoints are also shown in Figure 1.
The error bars are correlated at level of
$\sim 10 \%$. The $\sim 4 \%$ calibration error is not plotted in 
the figure. The beam/pointing  errors are of order of $\sim 10 \%$ at
$\ell =1000$ (see \cite{lee}).

{\bf Features in the CMB power spectrum.} 

Before going on to parameter extraction, 
it is important to try to quantify how well the present data provide 
evidence for multiple and coherent oscillations. 
Fits to the CMB data with phenomenological functions have been 
already extensively used in the past (see e.g. \cite{graca}, 
\cite{scott}, \cite{page}).
More recently, similar analyses have been carried out,
using parabolas (\cite{debe01}, \cite{bohdan}) 
or more elaborate oscillating functions with a well defined
frequency and phase (\cite{douspis}).

Since the first peak is evident, the statistical significance
of the secondary oscillations is now of greater interest.
In \cite{debe01} the BOOMERanG data bins centered 
at $450 < \ell < 1000$ were analyzed.
Using a Bayesian approach, a
 linear fit $C_\ell^T = C_A + C_B \ell$ is rejected at near 
$2 \sigma$ confidence level.
Also in \cite{debe01}, using a parabolic fit to the data, 
interleaved peaks and dips were found at $\ell =$ $215 \pm 11$, $431 \pm 10$,
$522 \pm 27$, $736 \pm 21$ and $837 \pm 15$ with 
amplitudes of the features $5760^{+344}_{-324}$, 
$1890^{+196}_{-178}$, $2290^{+330}_{-290}$,
$1640^{+500}_{-380}$, and $2210^{+900}_{-640}$ $\mu K^2$, correspondingly. 
The reported significance of the detection is $1.7\,\sigma$ for 
the second peak and dip, and $2.2\,\sigma$ for the third peak. 

The evidence for oscillations in the MAXIMA data 
has been carefully studied in \cite{stompor}.
While there is no evidence for a second peak, the power 
spectrum shows excess power at $\ell \sim 860$ over the average 
level of power at $411 \le\ell \le 785$ on the $95 \%$ confidence level.
Such a feature is consistent with the presence of a third acoustic peak.
         
In \cite{bohdan} the BOOMERanG, DASI and MAXIMA data were included in
a similar analysis. Both DASI and MAXIMA confirmed the main features
of the Boomerang CMB power spectrum: a dominant first acoustic peak
at $\ell \sim 200$, DASI shows a second peak at $\ell \sim 540$
and MAXIMA-I exhibits mainly a 'third peak' at $\ell \sim 840$.

Finally and more recently, in \cite{douspis} a different analysis 
was made, based on a function that smoothly interpolates between a spectrum 
with no oscillations and one with oscillations.
Again, within the context of this different phenomenological model,
a $2 \sigma$ presence for secondary oscillations was found.

\section{Consequences for Cosmology}

In principle, the CDM scenario of structure formation based on adiabatic
primordial fluctuations can depend on more than $11$ parameters.

However for a first analysis, it is useful to restrict ourselves to 
just $5$ parameters: the tilt of primordial spectrum of perturbations 
$n_S$, the optical depth of the universe $\tau_c$,
the density in baryons and dark matter
$\omega_b=\Omega_bh^2$ and $\omega_{dm}=\Omega_{dm}h^2$ and 
the shift parameter ${\cal R}$ which is related to the geometry of the universe
through (see \cite{efsbond}, \cite{melou}):

\be
{\cal R}=2 \sqrt{|\Omega_k| / \Omega_m} / \chi(y)
\ee

where $\Omega_m=\Omega_b+\Omega_{dm}$, $\Omega_k=1-\Omega_m-\Omega_{\Lambda}$,
the function $\chi(y)$ is $y$, $\sin(y)$ or $\sinh(y)$ for flat, closed and
open universes respectively and

\be
y=\sqrt{|\Omega_k|}\int_0^{z_{dec}}
{[\Omega_m(1+z)^3+\Omega_k(1+z)^2+\Omega_{\Lambda}]^{-1/2} dz}.
\ee

\begin{figure}[htb]
\begin{center}
\includegraphics[scale=0.38]{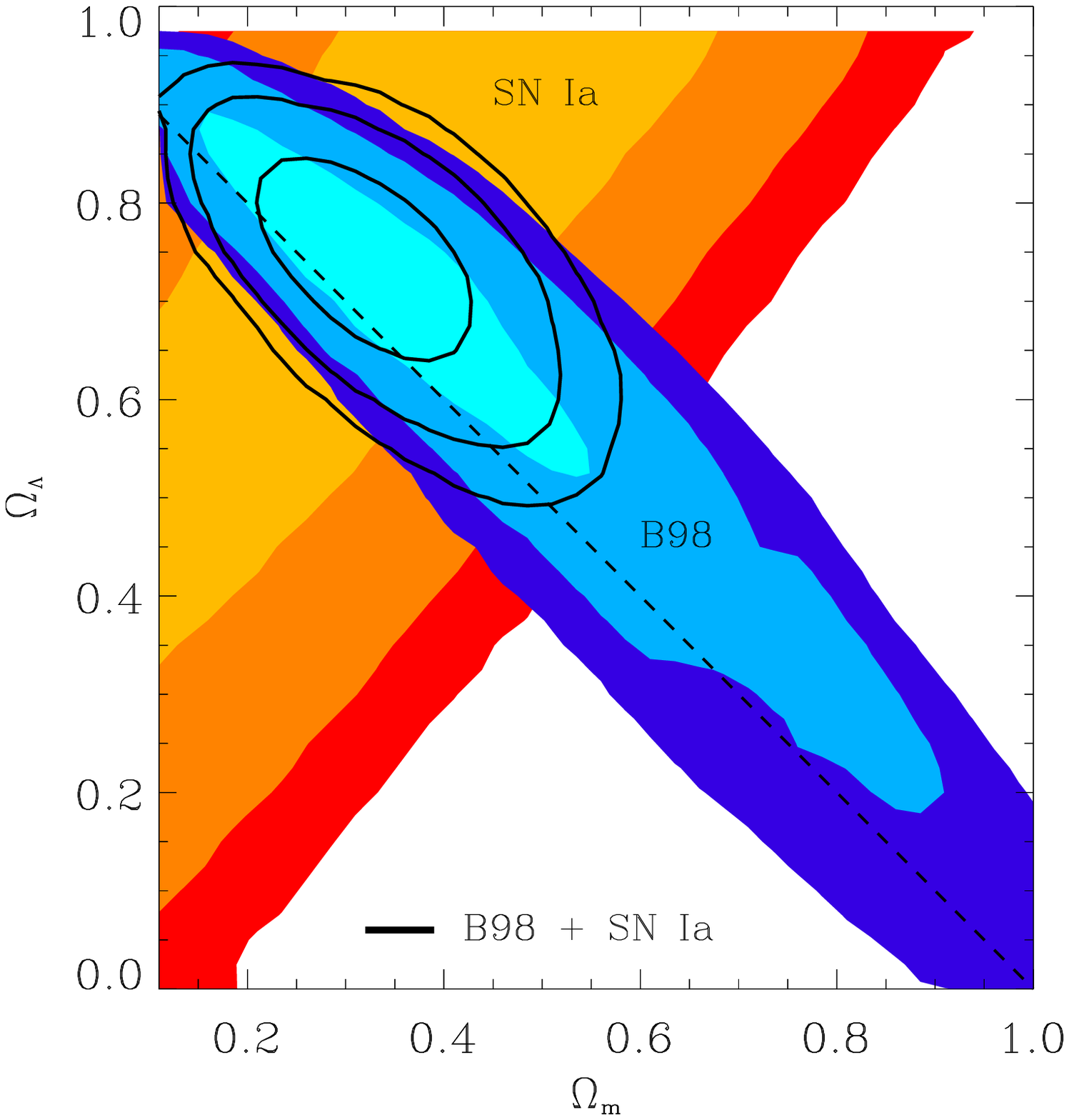}
\includegraphics[scale=0.38]{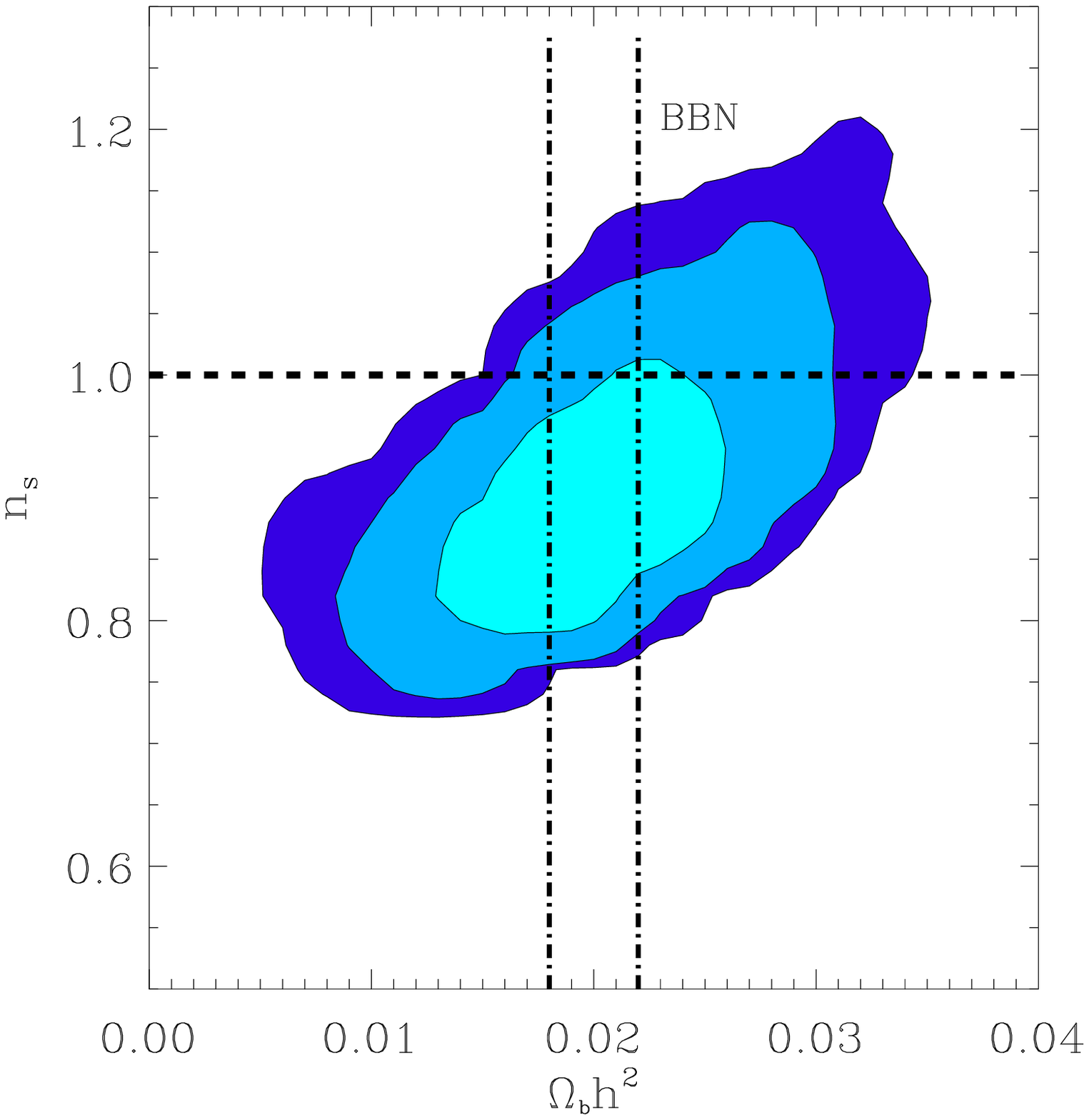}
\end{center}
\caption{Confidence contours in the $\Omega_M - \Omega_{\Lambda}$ 
and $\Omega_bh^2-n_S$ planes. Picture taken from \cite{debe01}.}
\label{fig3}
\end{figure}

The restriction of the analysis to only $5$ parameters is justified 
since a reasonable fit to the
data can be obtained with no additional parameters. 

In Fig. 2 we plot the likelihood contours on the 
$\Omega_{M}-\Omega_{\Lambda}$ and $\Omega_bh^2-n_S$ planes
from the BOOMERanG experiment as reported in \cite{debe01}.
Since the quantity ${\cal R}$ depends on $\Omega_{\Lambda}$ and 
$\Omega_{M}$ the CMB constraints on this parameter can be 
plotted on this plane.
As we can see from the top panel in the figure 
the data strongly suggest a flat universe
(i.e. $\Omega=\Omega_M+\Omega_{\Lambda}=1$). From the 
latest BOOMERanG data one obtains $\Omega=1.02\pm0.06$ (\cite{netterfield}).

The inclusion of complementary datasets in the analysis
breaks the angular diameter distance degeneracy in $\cal R$ 
and provides evidence
for a cosmological constant at high significance.
Adding the Hubble Space Telescope constraint on the Hubble
constant $h=0.72 \pm 0.08$ (\cite{freedman}, 
information from galaxy clustering and
from luminosity distance of type Ia supernovae 
gives (\cite{netterfield}) $\Omega_{\Lambda}=0.62_{-0.18}^{+0.10}$, 
$\Omega_{\Lambda}=0.55_{-0.09}^{+0.09}$ and 
$\Omega_{\Lambda}=0.73_{-0.07}^{+0.10}$ respectively.

Also interesting is the plot of the likelihood contours in the 
$\Omega_bh^2-n_S$ plane (Fig.2 bottom panel). 
As we can see, the present BOOMERanG data is in beautiful agreement with 
{\it both} a nearly scale invariant
spectrum of primordial fluctuations, as predicted by inflation, and
the value for the baryon density $\omega_b =0.020\pm0.002$ predicted
by Standard Big Bang Nucleosynthesis (see e.g. \cite{burles}).

An increase in the optical depth $\tau_c$ after recombination 
by reionization (see e.g. \cite{haiman} for a review) or by some more
exotic mechanism damps the amplitude of the CMB peaks.
Even if degeneracies with other parameters such as $n_S$ are present
(see e.g. \cite{debe97}) the BOOMERanG data provides the 
upper bound $\tau_c < 0.3$.

The amount of non-baryonic dark matter is also 
constrained by the CMB data with $\Omega_{dm}h^2=0.13 \pm 0.04$ 
at $68 \%$ c.l. (\cite{netterfield}).
The presence of power around the third peak is crucial in this sense,
since it cannot be easily accommodated in models based on just baryonic
matter (see e.g. \cite{bmp}, \cite{lmg}, \cite{mcgaugh} 
and references therein).

Furthermore, under the assumption of flatness, we can derive important
constraints on the age of the universe $t_0$ given by:

\begin{equation}
t_0=9.8 Gy \int_0^1{{a da \over
{[\omega_ma+\omega_{\Lambda}a^4]^{1/2}}}}
\end{equation}

In \cite{iggy} the BOOMERanG constraint on age has been compared 
with other independent results obtained from 
stellar populations in bright ellipticals,
 $^{238}$U age-measurement of an old halo star in our 
galaxy (\cite{cayrel}) and age the of the oldest halo globular cluster 
in the sample of Salaris \& Weiss (\cite{salaris}). 
All four methods give completely consistent results,
and enable us to set rigorous bounds on the maximum and minimum ages
that are allowed for the universe, $t_0=14 \pm 1$ GYrs (\cite{iggy},
\cite{netterfield},\cite{knoxage}).

The results from the DASI experiment have been extensively
reported in \cite{pryke} and are perfectly consistent with the
BOOMERanG results.
Pryke et al. report $\Omega=1.04 \pm 0.06$, $n_s=1.01^{0.08}_{0.06}$, 
$\Omega_bh^2=0.022^{0.004}_{0.003}$ and $\Omega_{dm}h^2=0.14 \pm 0.04$. 

The MAXIMA team reported similar compatible constraints in 
\cite{stompor}: $\Omega=0.9{+0.18\atop-0.16}$ and
$\Omega_b h^2=0.033{\pm 0.13}$ at $2 \sigma$ c.l..
However the MAXIMA data is not good enough to put strong 
constrains on the spectral index $n_S$ and the optical depth 
$\tau_c$ because of the degeneracy between the $2$ parameters.

\section{Open Questions.}

Even if the present CMB observations can be fitted with just $5$ 
parameters it is interesting to extend the analysis to other 
parameters allowed by the theory.
Here I will just summarize a few of them and discuss
 how well we can constrain them and what the effects 
on the results obtained in the previous section would be.

{\bf Gravity Waves.} 
The metric perturbations created during inflation belong to two types:
{\it scalar} perturbations, which couple to the stress-energy of 
matter in the universe and form the ``seeds'' for structure formation 
and {\it tensor} perturbations, also known as 
gravitational wave perturbations.
Both scalar and tensor perturbations contribute to CMB anisotropy.
In the recent CMB analysis by the BOOMERanG and DASI collaborations, 
the tensor modes have been neglected, 
even though a sizable background of gravity waves 
is expected in most of the inflationary scenarios. 
Furthermore, in the simplest models,
a detection of the GW background 
can provide information on the second derivative
of the inflaton potential and shed light on the physics at
$\sim 10^{16} Gev$ (see e.g. \cite{hoffman}).

The shape of the $C^T_{\ell}$ spectrum from tensor modes is drastically
different from the one expected from scalar fluctuations,
affecting only large angular scales (see e.g. \cite{crittenden}). 
The effect of including tensor modes is similar to 
just a rescaling of the degree-scale $COBE$ normalization and/or 
a removal of the corresponding data points from the analysis.

This further increases the degeneracies among cosmological
parameters, affecting mainly the estimates of the baryon and 
cold dark matter densities and the scalar spectral index $n_S$
(\cite{melk99},\cite{kmr}, \cite{wang}, \cite{efstathiougw}).

The amplitude of the GW background is therefore weakly constrained
by the CMB data alone, however, when information from BBN, local
cluster abundance and galaxy clustering are included, an upper limit
of about $r = C_2^T/C_2^S < 0.5$ is obtained.

{\bf Scale-dependence of the spectral index.}
The possibility of a scale dependence of the
scalar spectral index, $n_S(k)$, has been considered
in various works (see e.g. \cite{kosowsky}, \cite{copeland}, 
\cite{lythcovi}, \cite{doste}).
Even though this dependence is considered to 
have small effects on CMB scales in most of the slow-roll inflationary models, 
it is worthwhile to see if any useful constraint can be obtained.
Allowing the power spectrum to bend erases the ability 
of the CMB data to measure the tensor to scalar perturbation ratio
and enlarge the uncertainties on many cosmological parameters.

Recently, Covi and Lyth (\cite{covi}) investigated the
two-parameter scale-dependent spectral index
predicted by running-mass inflation models, and 
found that present CMB data allow for a significant scale-dependence of $n_S$.
In Hannestad et al. (\cite{hhv}, \cite{hhvh}) 
the case of a running spectral index has been studied, 
expanding the power spectrum $P(k)$ to second order in $ln(k)$. 
Again, their result indicates that a bend in the spectrum is
consistent with the CMB data.

Furthermore, phase transitions associated with spontaneous 
symmetry breaking during the inflationary era could result
in the breaking of the scale-invariance of the primordial density 
perturbation.
In \cite{sarkar}, \cite{louise} and \cite{ywang} 
the possibility of having step or bump-like features in the spectrum 
has also been considered.

While much of this work was motivated by the tension between the initial
release of the data and the baryonic abundance value from BBN, 
a sizable feature in the spectrum is still compatible with the latest
CMB data (\cite{elgaroy}).

{\bf Quintessence.}
The discovery that the universe's evolution may be dominated by
an effective cosmological constant \cite{super1}
is one of the most remarkable cosmological findings of recent years.
One candidate that could possibly explain the observations is a
dynamical scalar ``quintessence'' field. One of the strongest aspects of
quintessence theories is that they go some way towards explaining the
fine-tuning problem, that is why the energy density producing the acceleration
is $\sim 10^{-120}M_{pl}^{4}$. A vast range of ``tracker'' (see for
example \cite{quint,brax}) and ``scaling'' (for example
\cite{wett}, \cite{ferjoy}) quintessence models exist which approach
attractor solutions, giving the required energy density, independent
of initial conditions.
The common characteristic of quintessence
models is that their equations of state, $w_{Q}=p/\rho$, vary with time
while a cosmological constant remains fixed at
$w_{Q=\Lambda}=-1$ (see e.g. \cite{blud}). 
Observationally distinguishing a time variation in
the equation of state or finding $w_Q$ different from $-1$ will
therefore be a success for the quintessential scenario.
Quintessence can also affect the CMB by acting as an additional 
energy component with a characteristic viscosity.
However any early-universe imprint of quintessence 
is strongly constrained by Big Bang Nucleosynthesis
with $\Omega_Q(MeV) < 0.045$ at $2\sigma$ for temperatures near 
$T \sim 1Mev$ (\cite{bhm}, \cite{mathews}).

In \cite{rachel} we have combined the latest observations of the 
CMB anisotropies and the information from Large
Scale Structure (LSS) with the luminosity distance of high
redshift supernovae (SN-Ia) to put constraints on the dark energy
equation of state parameterized by a redshift independent
quintessence-field pressure-to-density ratio $w_Q$.                  
          
The importance of combining different data sets in order to obtain
reliable constraints on $w_Q$ has been stressed by
many authors (see e.g. \cite{PTW}, \cite{hugen},\cite{jochen}),
since each dataset suffers from degeneracies between the various
cosmological parameters and $w_Q$ . Even if one restricts consideration
 to flat universes and to a value of $w_Q$ constant
in time then the SN-Ia luminosity distance and position of the
first CMB peak are highly degenerate in $w_Q$ and $\Omega_Q$,
the energy density in quintessence.                              

\begin{figure}[htb]
\begin{center}
\includegraphics[scale=0.38]{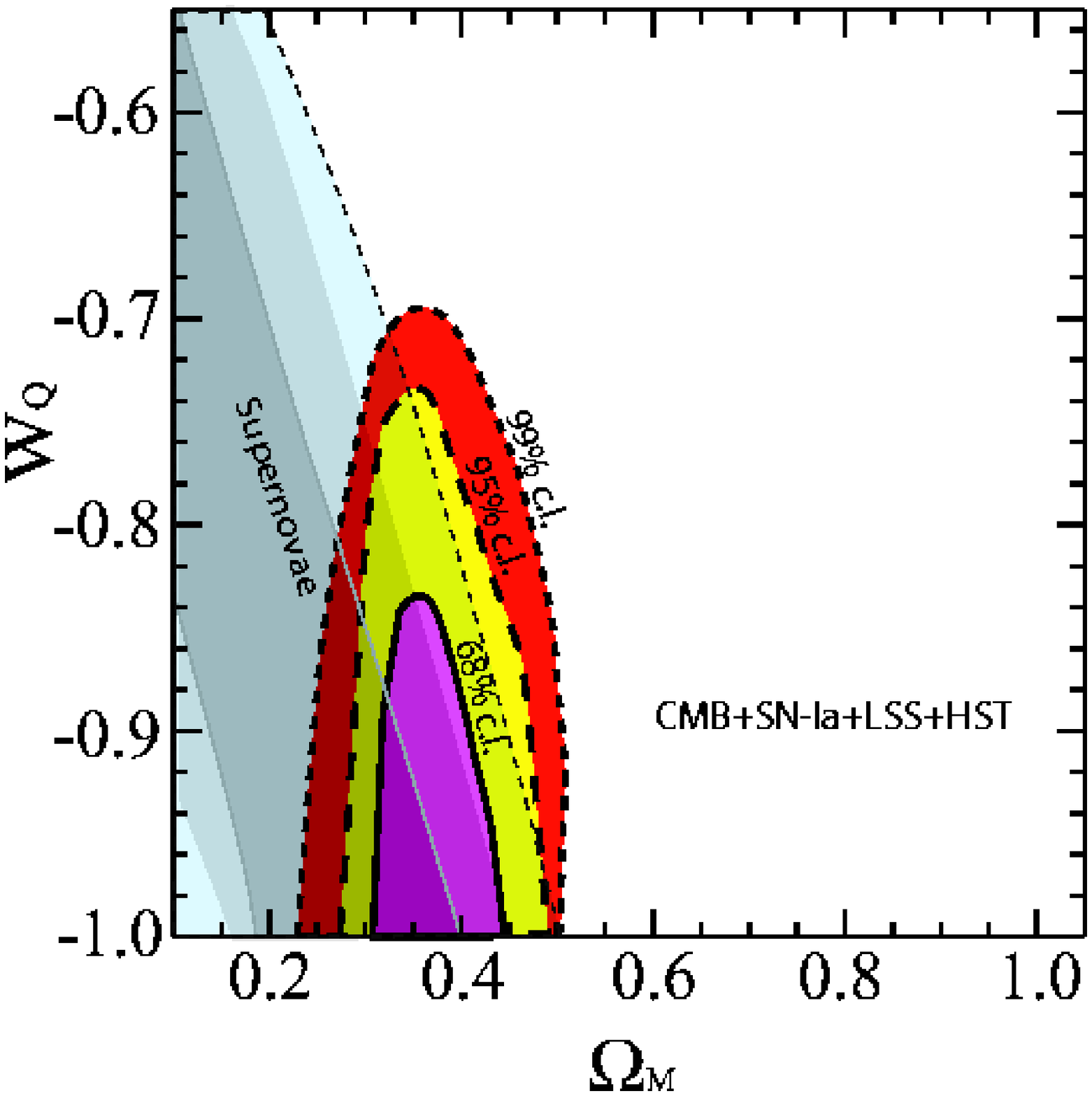}
\end{center}
\caption{The likelihood contours in the ($\Omega_M$, $w_Q$) plane,
with the remaining parameters taking their best-fitting values for the
joint CMB+SN-Ia+LSS analysis described in the text.
The contours correspond to 0.32, 0.05 and 0.01 of the peak value of the
likelihood, which are the 68\%, 95\% and 99\% confidence levels respectively.
Picture taken from \cite{rachel}.}
\label{figo1}
\end{figure}
\medskip                                                                      

In Figure 3 we plot the likelihood contours in the ($\Omega_M$, $w_Q$) plane 
for the joint analyses of CMB+SN-Ia+HST+LSS of \cite{rachel} together with 
the contours from the SN-Ia dataset only. 
As we can see, the combination of the datasets breaks the
luminosity distance degeneracy and suggests the presence of dark
energy with high significance. 
Furthermore, the new CMB results provided by Boomerang and DASI improve 
the constraints 
from previous and similar analysis (see e.g., \cite{PTW},\cite{bondq}), with
$w_Q<-0.85$ at $68 \%$ c.l..
Our final result is then perfectly in agreement with the $w_Q=-1$
cosmological constant case and gives no support to a
quintessential field scenario with $w_Q > -1$.

{\bf Big Bang Nucleosynthesis and Neutrinos.}
As we saw in the previous section, the SBBN $95 \%$ CL region,
corresponding to $\Omega_b h^2= 0.020 {\pm} 0.002$ ($95 \%$ c.l.), 
has a large overlap with the analogous CMBR contour. 
This fact, if it will be confirmed by future experiments on CMB
anisotropies, can be seen as one of the greatest
success, up to now, of the standard hot big bang model.

SBBN is well known to provide strong bounds on the number 
of relativistic species $N_\nu$. On the other hand,
Degenerate BBN (DBBN), first analyzed in Ref. \cite{d1,d2,d3,Kang}, gives
very weak constraint on the effective number of massless neutrinos, since
an increase in $N_\nu$ can be compensated by a change in both the chemical
potential of the electron neutrino, $\mu_{\nu_e}= \xi_e T$,
and $\Omega_bh^2$. 
Practically, SBBN relies on the theoretical assumption that 
background neutrinos have negligible chemical potential, just like their 
charged lepton partners. Even
though this hypothesis is perfectly justified by Occam razor, models have
been proposed in the literature
\cite{dibari,AF,DK,DolgovRep,Casas,MMR,McDonald,Foot}, where large neutrino
chemical potentials can be generated. It is therefore an interesting issue
for cosmology, as well as for our understanding of fundamental
interactions, to try to constrain the neutrino--antineutrino asymmetry
with the cosmological observables. It is well known that degenerate BBN gives
severe constraints on the electron neutrino chemical potential, $-0.06\leq
\xi_e\leq 1.1$, and weaker bounds on the chemical potentials of both 
the $\mu$ and $\tau$
neutrino, $|\xi_{\mu,\tau}|
\leq 5.6 \div 6.9$ \cite{Kang}, since electron neutrinos are directly
involved in neutron to proton conversion processes which eventually fix the
total amount of $^4He$ produced in nucleosynthesis, while $\xi_{\mu,\tau}$
only enters via their contribution to the expansion rate of the universe.

Combining the DBBN scenario with the bound on baryonic and radiation densities
allowed by CMBR data, it is possible to obtain strong constraints
on the parameters of the model. 
Such an analysis was previously performed in
(\cite{th7}, \cite{peloso}, \cite{hannestad}, \cite{orito}) 
using the first data release of BOOMERanG and MAXIMA 
(\cite{debe00}, \cite{hanany}). 
We recall that the neutrino chemical potentials contribute
to the total neutrino effective degrees of freedom $N_\nu$ as

\begin{equation}
N_{{\nu}} = 3 + \Sigma_{\alpha} \left[ \frac{30}{7}
\left( \frac{\xi_\alpha}{\pi} \right)^2 +
\frac{15}{7} \left( \frac{\xi_\alpha}{\pi} \right)^4 \right] \, .
\end{equation}

Notice that in order to get a bound on $\xi_\alpha$ we have here assumed   
that all relativistic degrees of freedom, other than photons, are given by
three (possibly) degenerate active neutrinos.

\begin{figure}[htb]
\begin{center}
\includegraphics[scale=0.4]{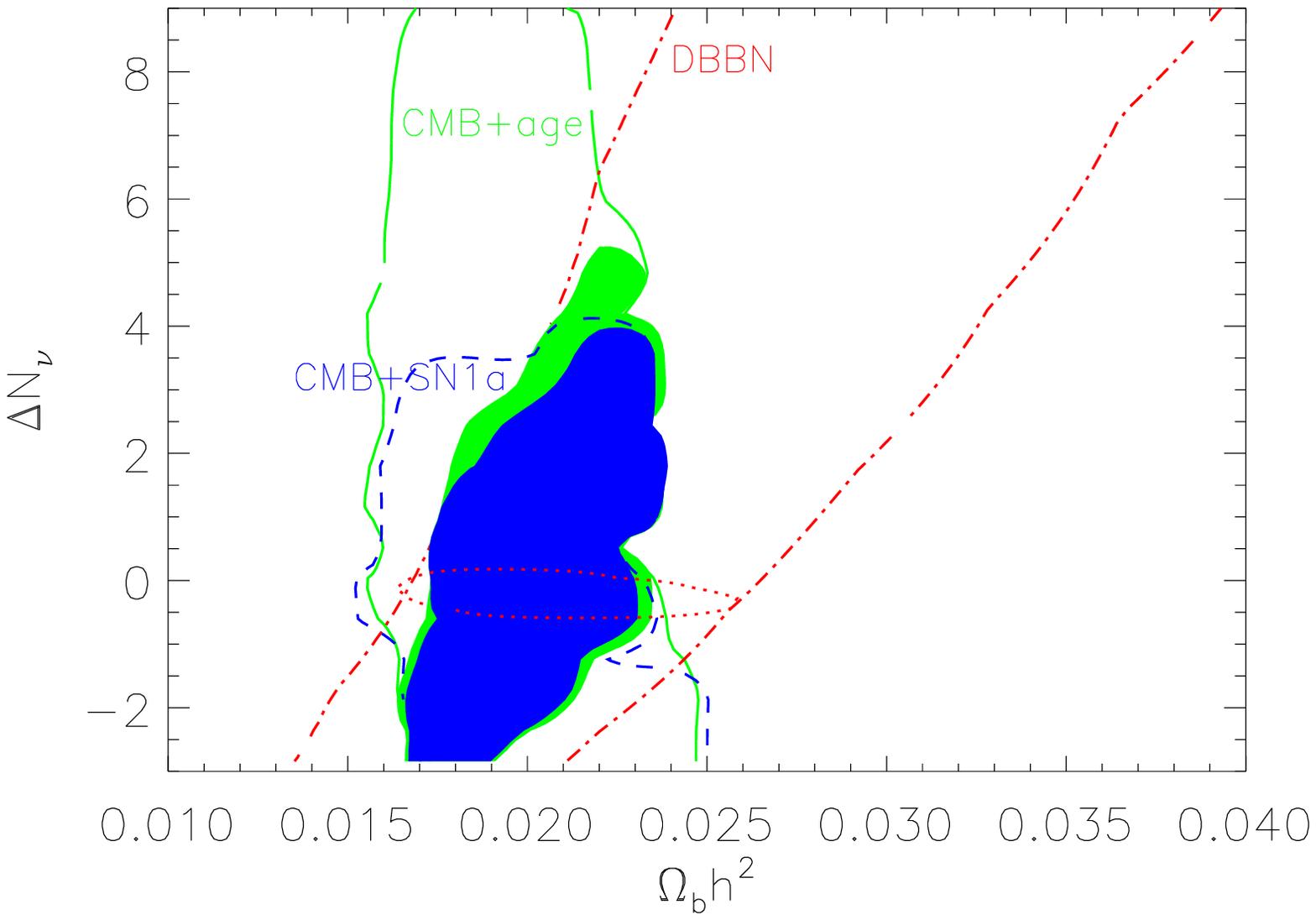}
\end{center}
\caption{The $95\%$ CL contours for degenerate BBN (dot-dashed (green) line),
new CMB results with just the age prior, $t>11$gyr (full (red) line),
and with just the SN1a prior (dashed (blue) line).
The combined analysis corresponds to the filled regions.
Marginalization leads to the bound
$\Omega_b h^2 =  0.020{\pm} 0.0035$ and
$N_\nu < 7$, both at $95 \%$, for DBBN+CMB+SN.
The dotted (green) line is the $95 \%$ CL allowed by SBBN.
Picture taken from \cite{hmmmp}.}
\label{figDbbn}
\end{figure}

Figure~4 summarizes the main results with the new CMB data, reported in
\cite{hmmmp} for the DBBN scenario. We plot the 
$95\%$ CL contours allowed by DBBN (dot-dashed (green) line),
together with the analogous $95 \%$ CL 
region coming from the CMB data analysis,
with only weak age prior, $t_0 > 11 $gyr (full (red) line).

Finally, the solid contour (light, red) is
the $95 \%$ CL region of the joint product distribution ${\cal L} \equiv
{\cal L}_{DBBN}$${\cdot}{\cal L}_{CMB}$.
The main new feature, with respect to the results
of Ref. \cite{th7} is that the resolution of the third peak shifts the CMB
likelihood contour towards smaller values for $\Omega_b h^2$, so when
combined with DBBN results, it singles out smaller values for $N_\nu$. In
fact from our analysis we get the bound $N_\nu \leq 8$, at $95 \%$ CL,
which translates into the new bounds $-0.01\leq \xi_e \leq 0.25$, and
$|\xi_{\mu,\tau}|\leq 2.9$, sensibly more stringent than what can be
found from DBBN alone.

A similar analysis can also be performed combining CMBR and DBBN data with
the Supernova Ia data \cite{super1}, which strongly reduces the degeneracy
between $\Omega_m$ and $\Omega_\Lambda$. At $95 \%$ C.L. we find
$\Delta N_\nu < 7 $, corresponding to  $-0.01\leq \xi_e \leq 0.22$ and
$|\xi_{\mu,\tau}|\leq 2.6$.  

Compatible results have been obtained in 
similar analyses (\cite{kneller},\cite{hannestad2}).

Some caution is naturally necessary when comparing the effective number of
neutrino degrees of freedom from BBN and CMB, since they may be related to
different physics. In fact the energy density in relativistic species may
change from the time of BBN ($T \sim MeV$) to the time of last rescattering 
($T\sim eV$).

{\bf Varying $\alpha$.} 
There are quite a large number of experimental constraints on the value of
fine structure constant $\alpha$. 
These measurements cover a wide range of timescales (see
\cite{alpharev} for a review of this subject), starting from present-day
laboratories ($z \sim 0$), geophysical tests ($z << 1$), and quasars
($z\sim 1 \div 3$), through the CMB ($z\sim 10^3$) and BBN ($z\sim10^{10}$)
bounds.

The recent analysis of \cite{Webb} of fine splitting of quasar doublet 
absorption lines gives a $4\sigma$ evidence for a time variation 
of $\alpha$, $\Delta \alpha/\alpha=(-0.72 {\pm} 0.18) 10^{-5}$, 
for the redshift range $z \sim 0.5 - 3.5$. 
This positive result was obtained using a many-multiplet
method, which, it is claimed, achieves an order of magnitude greater precision
than the alkali doublet method. Some of the initial ambiguities of the method
have been tackled by the authors with an improved technique, in which a
range of ions is considered, with varying dependence on $\alpha$, which
helps reduce possible problems such as varying isotope ratios, calibration
errors and possible Doppler shifts between different populations of ions
\cite{Quas1,Quas2,Quas3,Quas4}.

The present analysis of the $\alpha$-dependence relevant cosmological
observables like the anisotropy of CMB, Large Scale Structure
 and the light element primordial abundances does not support evidence 
for variations of the fine-structure constant 
(see \cite{avelino}, \cite{martins} and references therein).

{\bf Isocurvature modes.} Another key assumption is that the 
primordial fluctuations were adiabatic.
Adiabaticity is not a necessary consequence of inflation though and 
many inflationary models have been constructed where
isocurvature perturbations would have generically been concomitantly
produced (see e.g. \cite{langlois}, \cite{gordon}, \cite{bartolo}).

In a phenomenological approach one should consider the most
general primordial perturbation, introduced by \cite{kavi}, and
described by a $5X5$ symmetric matrix-valued generalization of
the power spectrum.
As showed by \cite{kavi}, the inclusion of isocurvature perturbations with 
auto and cross-correlations modes has dramatic
effects on standard parameter estimation with uncertainties 
becoming of order one.

Even assuming priors such as flatness, the inclusion of isocurvature
modes significantly enlarges our constraints on the baryon density
\cite{trotta} and the scalar spectral index \cite{amendola}.
Pure isocurvature perturbations are highly excluded by present CMB
data (\cite{enq}).

As we saw in the first section, it is also possible to have {\it active} and 
{\it decoherent} perturbations such as those produced by an inhomogeneously
distributed form of matter like topological defects.
Models based on global defects like cosmic strings and textures are
excluded at high significance by the present data (see e.g. \cite{DKMrep}).
However a mixture of adiabatic$+$defects is still compatible with the
observations (\cite{bouchet}, \cite{DKMrep}).
In principle, toy models based on {\it active} perturbations can
be constructed \cite{turok} that can mimic inflation and retain
a good agreement with observations \cite{dkm2}.

{\bf Secondary anisotropies.}
Secondary anisotropies can be generated due to photon interactions
with the matter potential wells (for example in the Rees-Sciama effect 
\cite{rees86} and lensing \cite{seljak96b}). The other secondary
anisotropies are induced by the interaction of CMB photons with free
electrons such as in the Sunyaev-Zel'dovich (SZ) effect
\cite[]{suniaev80}, the Ostriker-Vishniac (OV) effect
(\cite{ostriker86}, \cite{vishniac87}), and because of 
early inhomogeneous reionisation (IHR) 
(\cite{aghanim96},\cite{gruzinov98},\cite{knox98}).

\begin{figure}[htb]
\begin{center}
\epsfxsize=12cm
\epsfysize=10cm
\epsffile{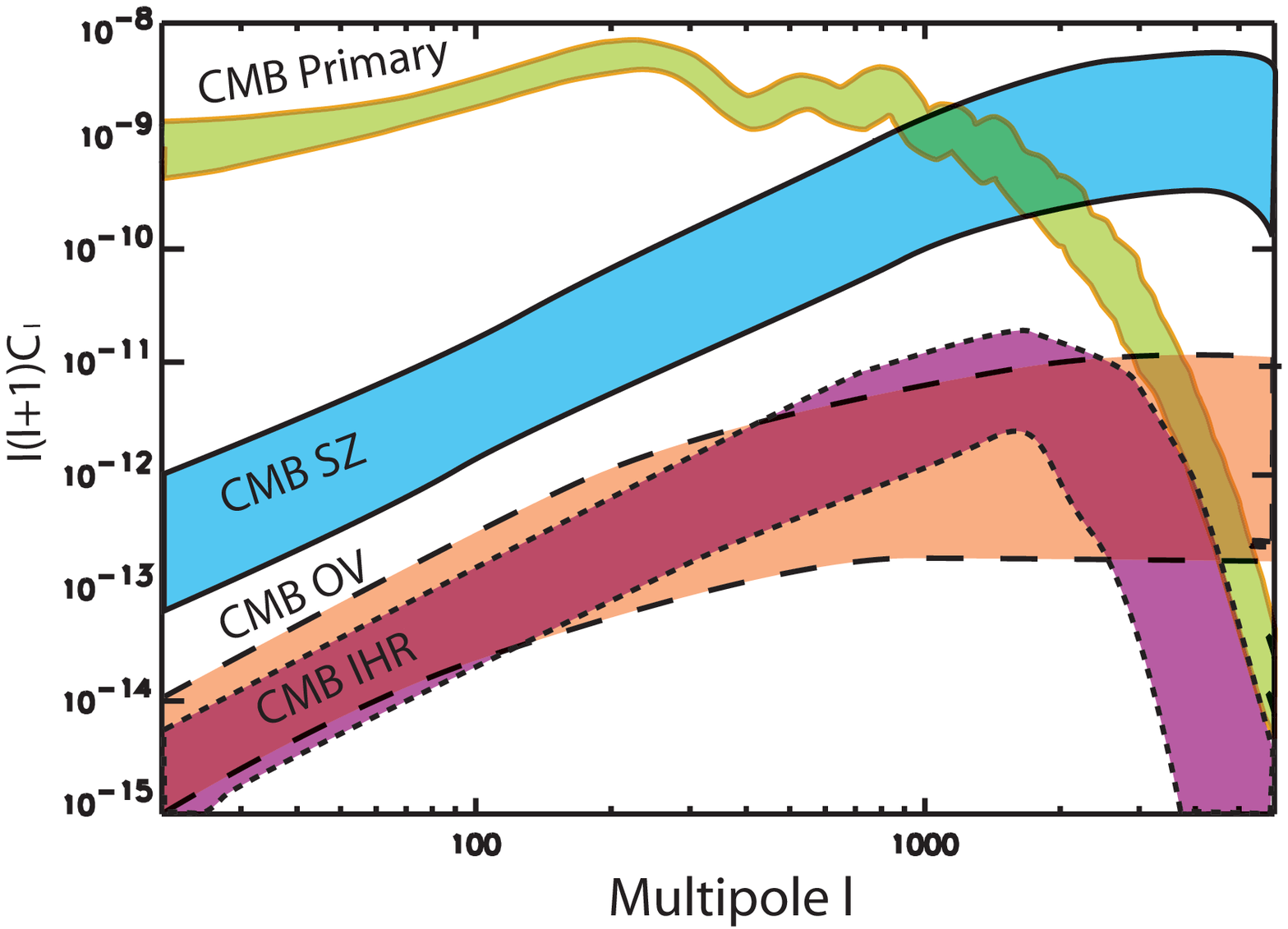}
\end{center}
\caption{Primary and Secondary CMB anisotropies.
The $95 \%$ confidence levels for the primary anisotropies are
plotted together with the extreme predictions allowed (within $2\sigma$)
by the present parameter estimations. The SZ effect at 150 GHz
is displayed in solid lines, the IHR and OV contributions are plotted
with dotted and dashed lines, respectively. }
\label{fig1}
\end{figure}
\medskip

Recent works (se e.g. \cite{asantha}, \cite{nabila} and references 
therein) have quantified the contribution of the secondary scattering
effects that are likely the dominant contributions at small scales.
In Figure 5, taken from \cite{nabila}, predictions for
the level of primary and secondary anisotropies are plotted, given the 
current status of observations. 
As we can see, in some extreme cosmological model, 
the secondary signal can be high enough to match the power of the 
CMB primary anisotropies after the third peak. 
Future small-scale CMB data such as that expected from
CBI \cite{mason2002}, will definitely be helpful in
scrutinising this results. The
measurements of the CMB anisotropies after the third peak will
therefore not only constrain the cosmological model through parameter 
estimation, but will also unable us to probe, via the secondary 
anisotropies (e.g. SZ), the formation and evolution of structures.

\section{Conclusions}

The recent CMB data represent a beautiful success for the 
standard cosmological model. The acoustic oscillations in
the CMB angular power spectrum, a major
prediction of the model, have now been detected at 
$\sim 5 \sigma$ C.L. for the first peak and $\sim 2 \sigma$ C.L.
for the second and third peak.
Furthermore, when constraints on cosmological parameters are 
derived under the assumption of adiabatic primordial perturbations
their values are in agreement with the predictions of the theory
and/or with independent observations.

As we saw in the previous section modifications 
as isocurvature modes or topological defects, are still
compatible with current CMB observations, but are not necessary and
can be reasonably constrained when complementary datasets are included.

Since the inflationary scenario is in agreement with the data and all 
the most relevant parameters are starting to be constrained within a 
few percent accuracy, the CMB is becoming a wonderful laboratory for 
investigating the possibilities of new physics. With the promise of 
large data sets from Map, Planck and SNAP satellites, 
opportunities may be open, for example, to constrain dark energy models, 
variations in fundamental constants and neutrino physics.

{\bf Acknowledgements}

I wish to thank the organizers of the conference: 
L. Celnikier and J. Tran ThanhVan.
Many thanks also to Nabila Aghanim, Rachel Bean, P.G. de Castro, 
Ruth Durrer, Steen Hansen, Pedro Ferreira, Will Kinney, Gianpiero Mangano, 
Carlos Martins, Gennaro Miele, Ofelia Pisanti, Antonio Riotto, 
Graca Rocha, Joe Silk, 
and Roberto Trotta for comments, discussions and help.

\end{document}